\documentclass[twocolumn,floatfix,prb,aps,showpacs,superscriptaddress,longbibliography]{revtex4-2}
\usepackage{amsmath, amssymb}

\usepackage{graphicx,amsmath,amssymb,color}
\usepackage{nicefrac}
\usepackage[titletoc,title]{appendix}
\usepackage{amsmath}
\usepackage{subfigure}

\usepackage[colorlinks,bookmarks=true,citecolor=blue,linkcolor=red,urlcolor=blue]{hyperref}
\usepackage[colorlinks,bookmarks=true,citecolor=blue,linkcolor=red,urlcolor=blue]{hyperref}
\usepackage{titlesec}

\makeatletter
\def\@fnsymbol#1{%
  \ifcase#1\relax 
  \or \ensuremath{\dagger}
  \or \ensuremath{*}
  \or \ensuremath{\mathsection}
  \or \ensuremath{\mathparagraph}
  \else\@ctrerr\fi
}
\makeatother

\begin{document}
\title{Spin liquid state in a three dimensional pyrochlore-like frustrated magnet}

\author{U. Jena}
\thanks{equal contribution}
    \affiliation{
    Department of Physics, Indian Institute of Technology Madras, Chennai, 600036, India}
    
    \author{S. Kundu}
    \thanks{equal contribution}
\affiliation{
Department of Physics, Indian Institute of Technology Madras, Chennai, 600036, India}

    \author{Suheon Lee}
\affiliation{
Center for Artificial Low Dimensional Electronic Systems, Institute for Basic Science, Pohang 37673, Republic of Korea}
 \author{Q. Faure}
\affiliation{
Laboratoire Léon Brillouin, CEA, CNRS, Université Paris-Saclay, CE-Saclay, F-91191 Gif-sur-Yvette, France}
\author{F. Damay}
\affiliation{
Laboratoire Léon Brillouin, CEA, CNRS, Université Paris-Saclay, CE-Saclay, F-91191 Gif-sur-Yvette, France}
\author{S. Rols}
\affiliation{
 Institut Laue-Langevin, 71 avenue des Martyrs, 38042 Grenoble Cedex 9, France}
\author{Adam Berlie}
\affiliation{
ISIS Neutron and Muon Source, STFC Rutherford Appleton Laboratory, Harwell Campus, Didcot, Oxfordshire OX11 0QX, United Kingdom}
\author{S. Petit}
\affiliation{
Laboratoire Léon Brillouin, CEA, CNRS, Université Paris-Saclay, CE-Saclay, F-91191 Gif-sur-Yvette, France}
\author{Kwang-Yong Choi}
\affiliation{
Department of Physics, Sungkyunkwan University, Suwon 16419, Republic of Korea}
\author{P. Khuntia}
\email{pkhuntia@iitm.ac.in}
\affiliation{
Department of Physics, Indian Institute of Technology Madras, Chennai, 600036, India}
\affiliation{
Quantum Centre of Excellence for Diamond and Emergent Materials,
Indian Institute of Technology Madras, Chennai, 600036, India}
	\date{\today}


	\begin{abstract}
The three-dimensional frustrated spin lattice in MgCrGaO\(_4\), where Cr\(^{3+}\) ions occupy a pyrochlore-like network, exemplifies a quantum magnet with competing interactions, macroscopic degeneracy, and exotic low‑energy excitations. Using thermodynamic, electron spin resonance (ESR), muon spin relaxation (\(\mu\)SR), and inelastic neutron scattering (INS) techniques, we observe no magnetic order or spin freezing down to 57 mK, despite a sizable exchange interaction (\(J = 58\,\text{K}\)) between Cr$^{3+}$ moments and inherent site disorder. Below the characteristic exchange energy scale, all probes detect the emergence of antiferromagnetic short‑range spin correlations, corroborated by magnetic diffuse scattering in the
wave vector dependence of low-energy magnetic excitations centered on \(Q = 1.5\,\text{\AA}^{-1}\) in inelastic neutron scattering experiments. The low‑temperature specific heat follows a near-quadratic dependence without a gap, consistent with algebraic spin correlations. These results establish MgCrGaO\(_4\) as a rare three‑dimensional classical spin liquid featuring a highly degenerate ground‑state manifold and gapless excitations, offering a strong impetus for the experimental realization of spin liquids
in higher-dimensional frustrated quantum magnets.

	\end{abstract}
	\maketitle

Highly frustrated magnets are potential contenders for hosting an array of exotic quantum and topological states with fractional quantum numbers, such as spin liquids with spinons, fractions, or Majorana fermions~\cite{balents2010spin, khuntia2020gapless, PhysRevLett.116.107203, savary2016quantum, kitaev2006anyons, knolle2019field}; while spin ices featuring monopole excitations coupled to emergent gauge fields are poised as some of the key attributes in contemporary condensed matter and its subfields~\cite{castelnovo2008magnetic, bramwell2001spin}. The magnetic behavior of 3$d$ ion-based pyrochlores is primarily
dictated by isotropic Heisenberg interactions, establishing
them as a prototypical platform for investigating emergent many-body phenomena, including classical spin liquids. Theoretically, the classical
Heisenberg antiferromagnet on a pyrochlore lattice with nearest-neighbor interactions is hypothesized
to exhibit a classical spin liquid state~\cite{PhysRevLett.80.2929}. Classical spin liquids are characterized by macroscopic ground state degeneracy~\cite{PhysRevLett.80.2929}, zero point entropy~\cite{ramirez1999zero}, non-quantum entanglement, and algebraic or topological spin correlations~\cite{henley2010coulomb, bramwell1993magnetization}. 

The experimental realization of spin liquids in three-dimensional spin lattices is uncommon, as the additional connectivity in the spin lattice and reduced quantum fluctuations in 3D quantum magnets frequently stabilize magnetic ordering or freezing, imposing strong constraints. A handful of 3D frustrated spin lattices have been reported as promising candidates for quantum spin liquids. Notably, the emblematic pyrochlore Ce$_2$T$_2$O$_7$ (\( T = \) Sn, Zr) features Ce$^{3+}$ ions with a \( J_{\text{eff}} = 1/2 \) Kramers doublet ground state arises from the interplay of spin-orbit coupling and the local \( D_{3d} \) crystal electric field, supporting a U(1) spin liquid with spinon excitations driven by frustrated magnetic octupolar interactions~\cite{PhysRevLett.115.097202, PhysRevB.95.041106, gao2019experimental, PhysRevX.12.021015}. Similarly, the spin-orbit-driven Mott insulator Na$_4$Ir$_3$O$_8$ features a corner-sharing network of triangular motifs with Ir$^{4+}$ (\( J_{\text{eff}} = 1/2 \)) ions on a 3D hyperkagome lattice exhibit a dynamical behavior even at elevated temperatures\cite{PhysRevLett.99.137207, PhysRevLett.101.197201, PhysRevLett.115.047201,PhysRevB.109.144405}.

PbCuTe\(_2\)O\(_6\), composed of corner-sharing triangles of Cu\(^{2+}\) (\(S = 1/2\)) forming a 3D hyperkagome lattice, displays hallmark signatures of a quantum spin liquid, including dispersive continua, short-range spin correlations, and a spinon Fermi surface~\cite{PhysRevLett.116.107203,PhysRevB.90.035141,chillal2020evidence,PhysRevLett.131.256701}. Cr-based pyrochlore CsNiCrF\(_6\) offers a viable platform to explore rich quantum phases, namely, transitions from Heisenberg models to Coulomb phases, driven by correlated disorder between Ni\(^{2+}\) and Cr\(^{3+}\) cations~\cite{fennell2019multiple}. This disorder destabilizes magnetic order, giving rise to gauge-constrained quantum phases with pinch points and unconventional low-energy excitations. Likewise, disordered \(S = 1\) pyrochlore NaCaNi\(_2\)F\(_7\) hosts a continuum of exotic magnetic excitations, with low-energy pinch points suggesting a Coulomb-like phase~\cite{plumb2019continuum}.  

Chromium-based spinels (ACr$_2$O$_4$, A = Zn$^{2+}$, Mg$^{2+}$, Cd$^{2+}$), in which the Cr$^{3+}$ ($S=3/2$) ions constitute a pyrochlore lattice with corner shared tetrahedra by occupying octahedral sites, resulting
in inherent magnetic frustration while the A$^{2+}$
ions are located in tetrahedral coordination. These 3D quantum magnets entail dominant antiferromagnetic exchange interaction with strong frustration manifested by large Curie-Weiss temperature compared to the magnetic ordering temperature ~\cite{PhysRevLett.94.047202,ortega2008low,tsurkan2021complexity}. ACr$_2$O$_4$ host a range of intriguing correlated phenomena, including spin-Peierls instability, spin liquids with exotic excitations, magnetic superfluidity, and complex magnetic ordering that are driven by geometric frustration, magnetostructural effects, competition between nearly degenerate states, and non-trivial spin fluctuation mechanisms~\cite{PhysRevLett.94.047202,ortega2008low,tsurkan2021complexity,PhysRevLett.110.097203,Lee_2007,PhysRevLett.134.086702,lee2002emergent,PhysRevLett.88.067203,PhysRevLett.122.097201,PhysRevB.87.214424}. The physics of chromium spinels is quite rich and intricate that provides
a plausible platform for exploring emergent low temperature many-body phenomena in pristine and disordered 3D frustrated spinels hosting a pyrochlore structure~\cite{tsurkan2021complexity, lee2002emergent}. Correlated disorder in frustrated
magnets induced by unavoidable anti-site disorder,
bond or exchange randomness is often unavoidable; however, it can serve as a
new prism to uncover many emergent quantum phases~\cite{gomilvsek2019kondo, RevModPhys.81.45, keen2015crystallography, kimchi2018scaling, khatua2022signature, jena2025nature}. Understanding the role of intrinsic disorder, chemical pressure, magnetostructural effects and interplay between competing degrees of freedom on the underlying ground state in 3D magnets is pivotal for the design of promising quantum magnets with the potential to host exotic quantum states.

 Herein, we present thermodynamic, ESR, $\mu$SR, and INS results on a three-dimensional disordered pyrochlore-like spin lattice MgCrGaO$_4$ (MCGO). Despite the three-dimensional nature of the spin lattice, MCGO does not undergo a phase transition down to 57 mK. $\mu$SR experiments rule out the presence of spin freezing down to 80 mK, which supports a spin liquid behavior. The specific heat shows a broad peak around 5 K accompanied by reduced entropy release, implying short-range spin correlations that are supported by ESR and $\mu$SR results. The low-energy inelastic neutron scattering adds further credence to the development of short-range antiferromagnetic dynamical spin correlations below 20 K, which is manifested by the presence of broad magnetic diffuse scattering centered on $Q = 1.5$\,\AA$^{-1}$.
The power-law behavior of specific heat below the broad maximum reflects highly degenerate low-energy excitations with algebraic spin correlations consistent with INS experiment in this 3D classical spin liquid candidate with spin $S>1/2$.
\begin{figure}[htpb]
		\begin{center}
			\includegraphics[width=0.5\textwidth,height=0.43
            \textwidth]{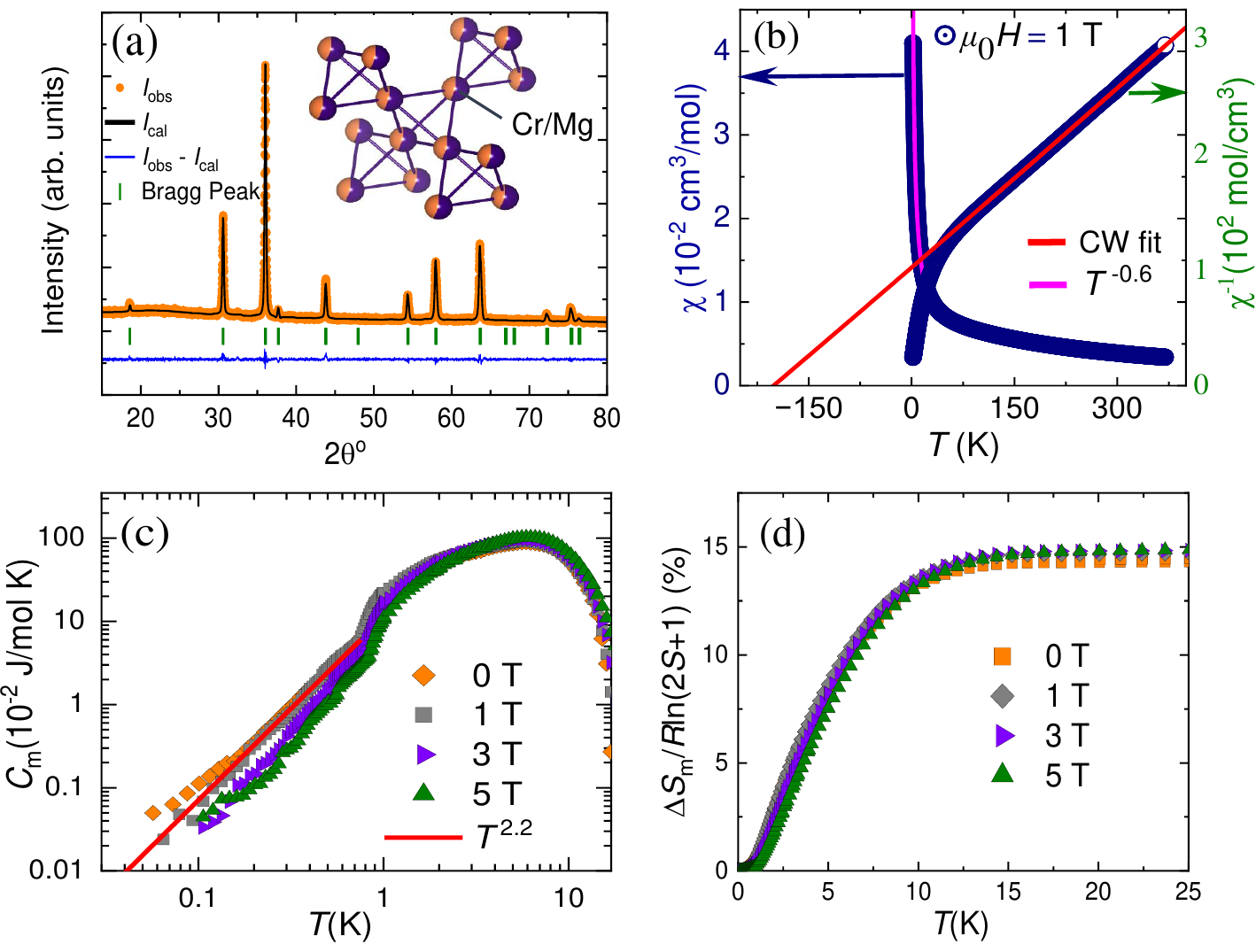}
			\caption{(a) The Rietveld refinement of the powder X-ray diffraction pattern of MgCrGaO$_4$. The inset shows the pyrochlore lattice formed by Cr$^{3+}$ moments in the crystal structure of MCGO, illustrating the geometric frustration inherent to the lattice. (b) Temperature-dependent magnetic susceptibility measured under an applied field of 1\,T. The inverse susceptibility is fitted to the Curie–Weiss law (solid red line), indicating dominant magnetic interactions. At low temperatures, the dc magnetic susceptibility follows a power-law dependence, $\chi(T) \sim T^{-0.6}$. (c) The magnetic heat capacity $C_\text{m}$ is plotted as a function of temperature in different fields. The $C_\text{m}$ data follow a power-law dependence, $C_\text{p} \propto T^{2.2}$, at low temperatures. (d) The magnetic entropy change $\Delta S_\text{m}$ is just $15\%$ of the expected theoretical values Rln 4 for one Cr$^{3+}$ ion per formula unit.
}
			\label{fig: Magnetization}
		\end{center}
	\end{figure}

 The Rietveld refinement of XRD data reveals [see
Fig. S1 in supplementary [SM]~\cite{supplement}] that MCGO adopts a spinel configuration (AB$_2$O$_4$) and crystallizes in a cubic structure with space group \textit{Fd}$\bar{3}m$ and lattice parameter $a=8.268$\,\AA. MCGO is
structurally analogous to
ZnCr$_{2x}$Ga$_{2-2x}$O$_4$ but with Mg replacing Zn. Specifically,
MCGO composition ($x = 0.5$) places it near the critical
doping region~\cite{fiorani1983magnetic}. The A-site comprises predominantly non-magnetic Ga$^{3+}$ and a fraction of Mg$^{2+}$ ions, while the B-site contains a mixed population of magnetic Cr$^{3+}$(56\%) ions and non-magnetic Mg$^{2+}$(44\%) ions, forming a pyrochlore-like three dimensional spin-lattice that accounts for geometric frustration (Fig.~\ref{fig: Magnetization}(a)). This unavoidable cationic site disorder is related to the intrinsic inversion tendency of MgGa$_2$O$_4$, a prototypical inverse spinel wherein Ga$^{3+}$ and Mg$^{2+}$ exhibit substantial A/B site mixing~\cite{huber1960repartition}. Structurally, the octahedral environment of O$^{2-}$ ions surrounding the B-site splits the $d$-orbitals of Cr$^{3+}$ into a lower-energy triply degenerate $t_{2g}$ orbital and a higher-energy doubly degenerate $e_{g}$ orbital~\cite{abragam2012electron, bethe1929termaufspaltung}. The electronic ground state of Cr$^{3+}$ (3$d^3$, $S=3/2$) resides within the $t_{2g}$ manifold, resulting in localized magnetic moments. In a pristine  Cr-based pyrochlore oxide, the order-by-disorder mechanism
lifts the classical degeneracy, selecting a unique
magnetic ground state. It is interesting to investigate the effect of inherent disorder on the ground state in the present 3D geometrically frustrated magnet.

The Curie-Weiss fit of the magnetic susceptibility, $\chi (T)$, of MCGO recorded in an applied magnetic field of 1 T (Fig.~\ref{fig: Magnetization}(b)) in the temperature range of 100–350 K yields a Curie-Weiss temperature of $\theta_{\text{CW}} = -201$ K, which indicates predominant antiferromagnetic exchange interaction between Cr$^{3+}$ moments, and an effective magnetic moment \(\mu_\text{eff} = 3.98 \, \mu_\text{B}\), slightly exceeding the spin-only value of \(3.87 \, \mu_\text{B}\) for \(S = 3/2\) Cr\(^{3+}\) ions. The significant reduction from $\theta_{\text{CW}} = -433$ K in MgCr$_2$O$_4$~\cite{PhysRevB.83.064407} to $-201$ K in MCGO suggests a weakening of antiferromagnetic exchange interaction, owing to inherent site disorder between Ga$^{3+}$ and  Cr$^{3+}$ that disrupts the magnetic interaction and reduces the extent of geometric frustration within the Cr$^{3+}$ sublattice. The deviation from Curie–Weiss behavior in the DC $\chi(T)$ below 50\,K suggests the emergence of locally entangled spin clusters that reduce the net moment through antiferromagnetic correlations. In the presence of exchange randomness and geometric frustration, such spin clusters can behave analogously to singlet states, which is further corroborated by ESR measurements (see next section). These local correlations effectively suppress uniform magnetization while preserving dynamic fluctuations, pointing to a complex magnetic ground state shaped by both disorder and frustration. This behavior reflects the intrinsic tendency of frustrated systems to form nontrivial, spatially inhomogeneous magnetic textures in the absence of long-range order~\cite{PhysRevB.22.1305, khatua2022signature}. The $\chi(T)$ recorded at 100~Oe shows no indication of spin freezing, as evidenced by the lack of bifurcation between zero-field-cooled (ZFC) and field-cooled (FC) curves (see SM\cite{supplement}) despite substantial anti-site disorder. Furthermore, the absence of hysteresis signature in isothermal magnetization down to 2 K (see Fig. S1~\cite{supplement}) adds further credence to the lack of spin-freezing or impurities in MCGO.

The estimation of exchange energy \( J \) for the current disordered frustrated spin-lattice hosting isotropic nearest neighbor exchange interactions is crucial for comprehending the dynamic spin correlations.  The nearest-neighbor exchange coupling was estimated to be $J = 58$ K, using a simple mean field approximation followed by spin wave formalism to extract the exchange interaction and the corresponding INS spectra, as discussed in the respective section (see ref.~\cite{supplement}). The spin stiffness in the large $S$ limit, \(\rho_s \sim \frac{J S^2}{d}\), where \(d\) represents the nearest-neighbor distance between Cr\(^{3+}\) ions in a tetrahedron, quantifies the energy cost associated with spin distortions~\cite{kittel1963quantum, auerbach2012interacting}. Using \(d = 2.923 \, \text{\AA}\), \(\rho_s \) is estimated as \(\sim 6.4 \, \text{meV/\AA}\), indicates relatively robust antiferromagnetic exchange interactions, consistent with the development of short-range spin correlations observed in this frustrated magnet~\cite{PhysRevB.48.13170}. The theoretical upper limit for the propagation speed of conventional magnetic excitations, \(c = \frac{2\sqrt{\text{D}} J S d}{\hbar}\), where \(\text{D} = 3\) represents the spin-lattice dimensionality, is calculated to be \(c \sim 11.1 \times 10^3 \, \text{m/s}\)~\cite{auerbach2012interacting}. While the inevitable site dilution remains above the percolation threshold (\(p_c \sim 0.39\) for a pyrochlore lattice)~\cite{henley2001effective, PhysRev.133.A310}, disorder softens the spin stiffness, according to the relation \(\rho_s(p) \sim \rho_{s,0} (p - p_c)^t\), where \(t\) is the critical exponent~\cite{stauffer2018introduction}. The diminishing stiffness and frustration-induced spin fluctuations weaken the system's ability to stabilize magnetic order even in three-dimensional lattices. While long wavelength fluctuations in 3D systems are less pronounced compared to the logarithmic divergences characteristic of 2D systems, they nonetheless exert a substantial influence on the magnetic properties~\cite{PhysRevB.39.2344, illing2017mermin}. In particular, the absence of glass transition despite unavoidable site disorder, combined with algebraic spin correlations and broadband relaxation spectra, points to an intricate interplay between correlated disorder and frustration~\cite{PhysRevB.98.134427, PhysRevX.8.031028}.

\begin{figure}[h]
	\centering
	\includegraphics[width=0.45\textwidth,height=0.38\textwidth]{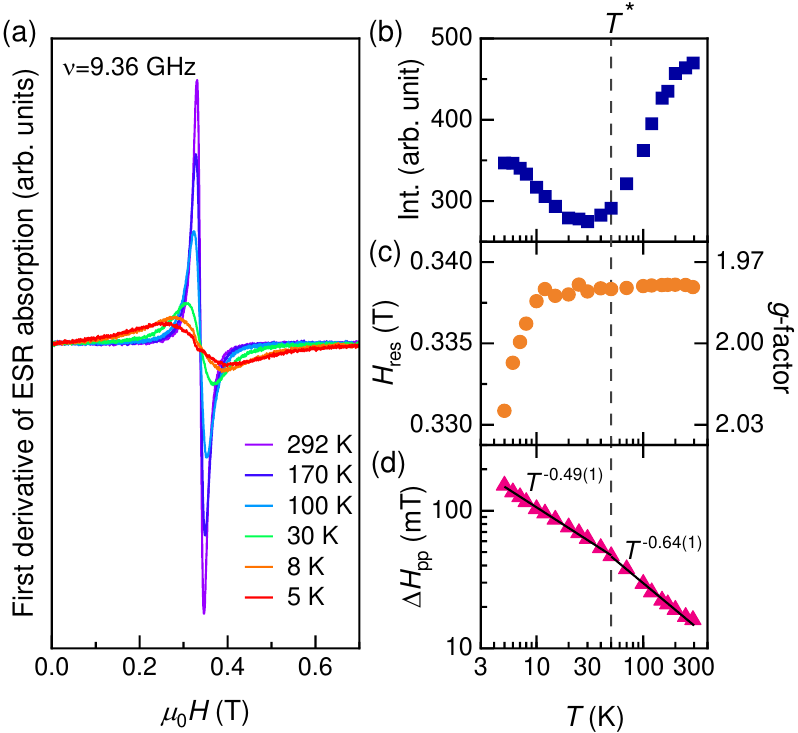}
	\caption{(a) X-band ESR spectra at various temperatures. (b) Temperature dependence of the integrated ESR intensity. (c) Resonance field as a function of temperature. (d) Peak-to-peak ESR linewidth vs. temperature. The solid lines denote the power-law dependence $\Delta H_\textrm{pp}\propto T^{-n}$. The dashed vertical line in (b)-(d) indicates the characteristic temperature $T^*=50$~K.}
	\label{Figure_ESR}
\end{figure}

Figure~\ref{Figure_ESR}(a) presents the X-band ESR spectra at selected temperatures. Upon cooling from 292~K, the ESR spectra gradually broaden while shifting to lower magnetic fields. Across the entire temperature range, the ESR signals are well described by a single Lorentzian profile. The extracted parameters—the integrated ESR intensity, the resonance field ($H_\textrm{res}$), and the peak-to-peak ESR linewidth ($\Delta H_\textrm{pp}$)-are shown in Figs.~\ref{Figure_ESR}(b)-(d). As the temperature is lowered, the ESR intensity gradually decreases, remains nearly constant below $T^*=50$~K, and exhibits an upturn below 20~K. Since the integrated ESR intensity is proportional to the concentration of unpaired electrons, this serves as the intrinsic spin susceptibility. Compared to the static susceptibility $\chi(T)$, the suppressed ESR intensity at high $T$s suggests a reduction in the number of unpaired electrons involved in magnetic resonance, likely due to singlet spin correlations originating from fragmented spin clusters. On the other hand, the enhancement of ESR intensity at low $T$ can be attributed to the development of magnetic correlations between Cr$^{3+}$ moments in fragmented spin clusters (see below). Note that we detect no impurity signals as there is no sharp peak near the $g$-factor of 2 ($\sim0.335$~T).

Next, we examine the temperature dependence of the resonance field and peak-to-peak linewidth. At high temperatures, $H_\textrm{res}$ remains nearly temperature-independent. However, as the temperature decreases through $T^*$, $H_\textrm{res}$ undergoes a subtle reduction, followed by a steep decrease below 10~K. As the field shift of $H_\textrm{res}$ reflects the development of internal magnetic fields, its gradual decrease below $T^*$ suggests the onset of short-range spin correlations. Moreover, the sharp reduction below 15~K coincides with the temperature range in which the dc magnetic susceptibility follows a power-law dependence, $\chi(T)\sim T^{-0.6}$ and with the exchange coupling constant $J=58$~K. This suggests that, at high temperatures, short-range spin correlations compete with thermal energy, preventing the establishment of internal fields despite the onset of spin correlations. However, at low temperatures, the exchange interactions dominate over thermal energy, leading to a correlated magnetic state below 15~K. Furthermore, $\Delta H_\textrm{pp}$ exhibits two distinct power-law behaviors: $T^{-0.64(1)}$ for $T>T^*$ and $T^{-0.49(1)}$for $T<T^*$, further supporting the change of spin correlations at $T^*$.

The specific heat measurements down to 57 mK in magnetic fields ranging from 0 to 5 T reveal crucial insights into the low-energy excitations of MCGO. No signature of a $\lambda$-type peak, typical of long-range magnetic ordering, is observed down to 57 mK. The obtained frustration parameter, defined as \( f = |\theta_\text{CW}| / T_N \sim 3500 \), suggests that strong frustration induced spin fluctuations and persistence of macroscopic ground state degeneracy, likely destabilize conventional long-range order. The magnetic specific heat ($C_\text{m}$) was obtained after subtracting the lattice specific heat using the Debye-Einstein model due to the unavailability of a non-magnetic analog (see SM\cite{supplement}). The emergence of a broad peak near 5~K in $C_\text{m}$ (Fig.~\ref{fig: Magnetization} (c)), indicates persistent short-range magnetic correlations at low temperatures, which is further substantiated by the release of only 15\% of the expected magnetic entropy compared to Rln4  for Cr$^{3+}$ ($S=3/2$) (see Fig. ~\ref{fig: Magnetization}(d)). Magnetic specific heat, \( C_\text{m} \) follows a power-law behavior (\( C_\text{m} \propto T^{2.2} \)) below 1 K (see Fig.~\ref{fig: Magnetization}(c)), which suggests gapless excitations and non-trivial spin dynamics characteristic of frustrated magnets hosting a classical spin liquid with algebraic spin correlations and a highly degenerate manifold of low energy excitations~\cite{khatua2023experimental}.


\begin{figure}[htbp]
    \centering
    \includegraphics[width=0.47\textwidth,height=0.28\textwidth]{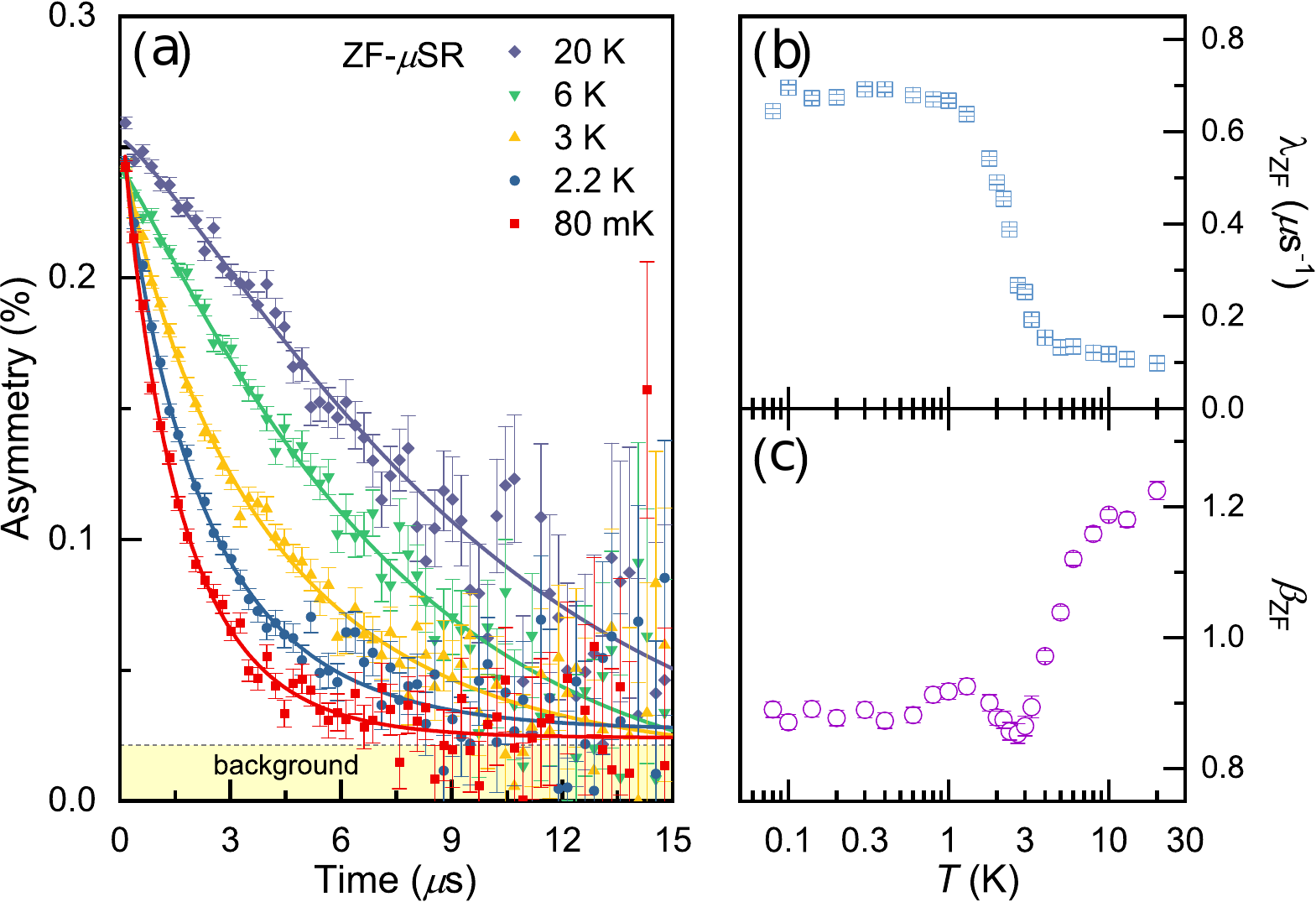} 
    \includegraphics[width=0.47\textwidth,height=0.28\textwidth]{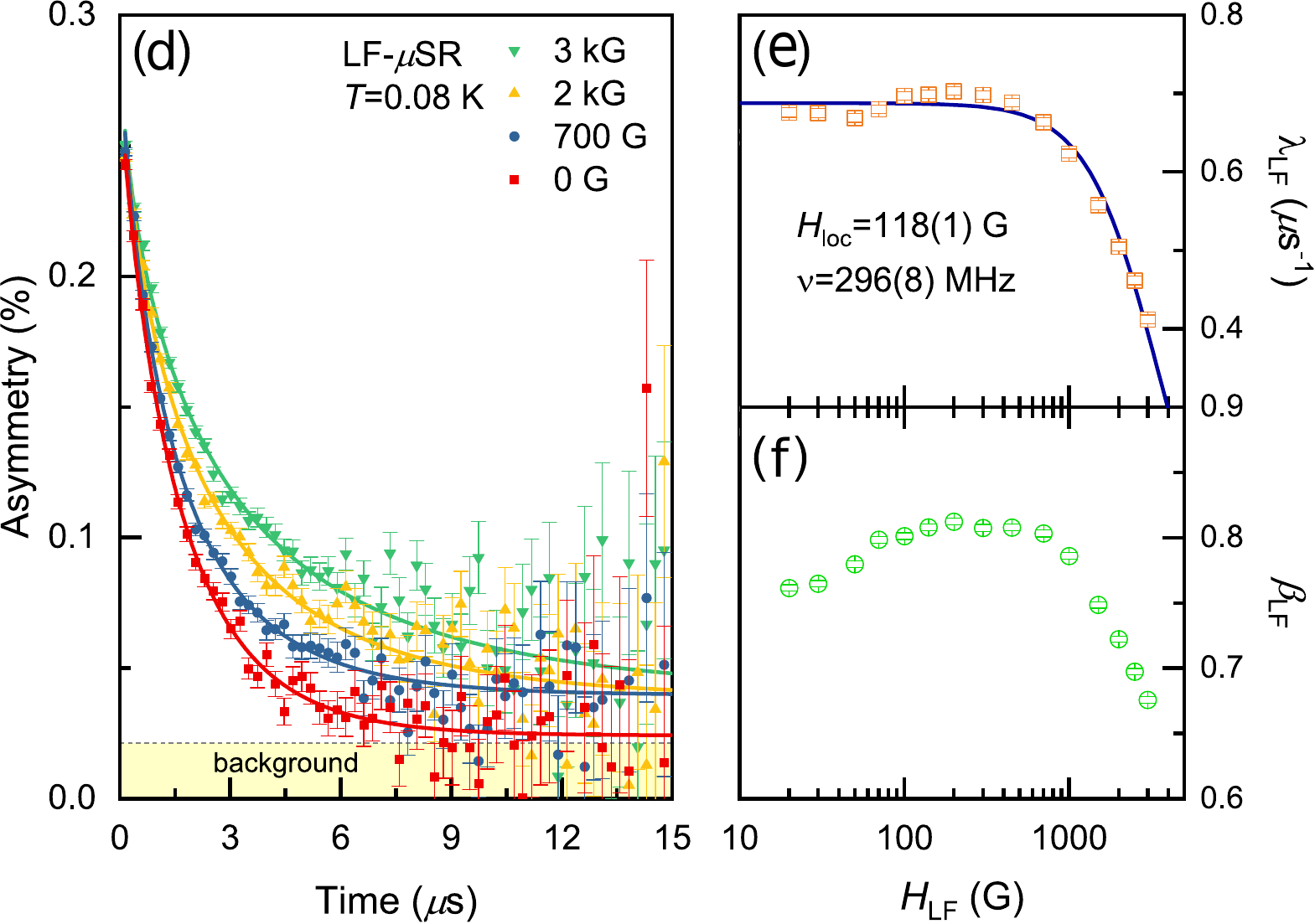} 
    \caption{(a) Representative ZF-$\mu$SR spectra at various temperatures. The solid curves indicate the fittings to stretched exponential. (b) Temperature dependence of the muon spin relaxation rate. (c) Stretched exponent as a function of temperature. (d) Longitudinal field dependence of the $\mu$SR spectra at $T=0.08$~K. The solid lines denote the stretched exponential fits as discussed in the text. (e) Muon spin relaxation rate as a function of longitudinal field. The solid curve represents the fits with the Redfield formula as discussed in the text. (f) Longitudinal field dependence of the stretched exponent.}
    \label{fig:stacked}
\end{figure}

To understand the spin dynamics in the ground state, we performed $\mu$SR measurements on MCGO. As depicted in Fig.~\ref{fig:stacked}(a), ZF-$\mu$SR asymmetries at high temperatures show an exponential decay in time. With decreasing temperature, the muon spin depolarization becomes gradually faster. We find no evidence for internal static magnetic fields down to 80~mK, including coherent oscillating signals and initial asymmetry loss. These observations suggest the absence of static magnetism in this frustrated magnet MCGO that is consistent with thermodynamic results. Furthermore, the absence of the so-called 1/3 tail in the muon asymmetry rules out any spin-freezing behavior down to 80 mK in this material.

\begin{figure*}[ht]
\includegraphics[width=1\textwidth]{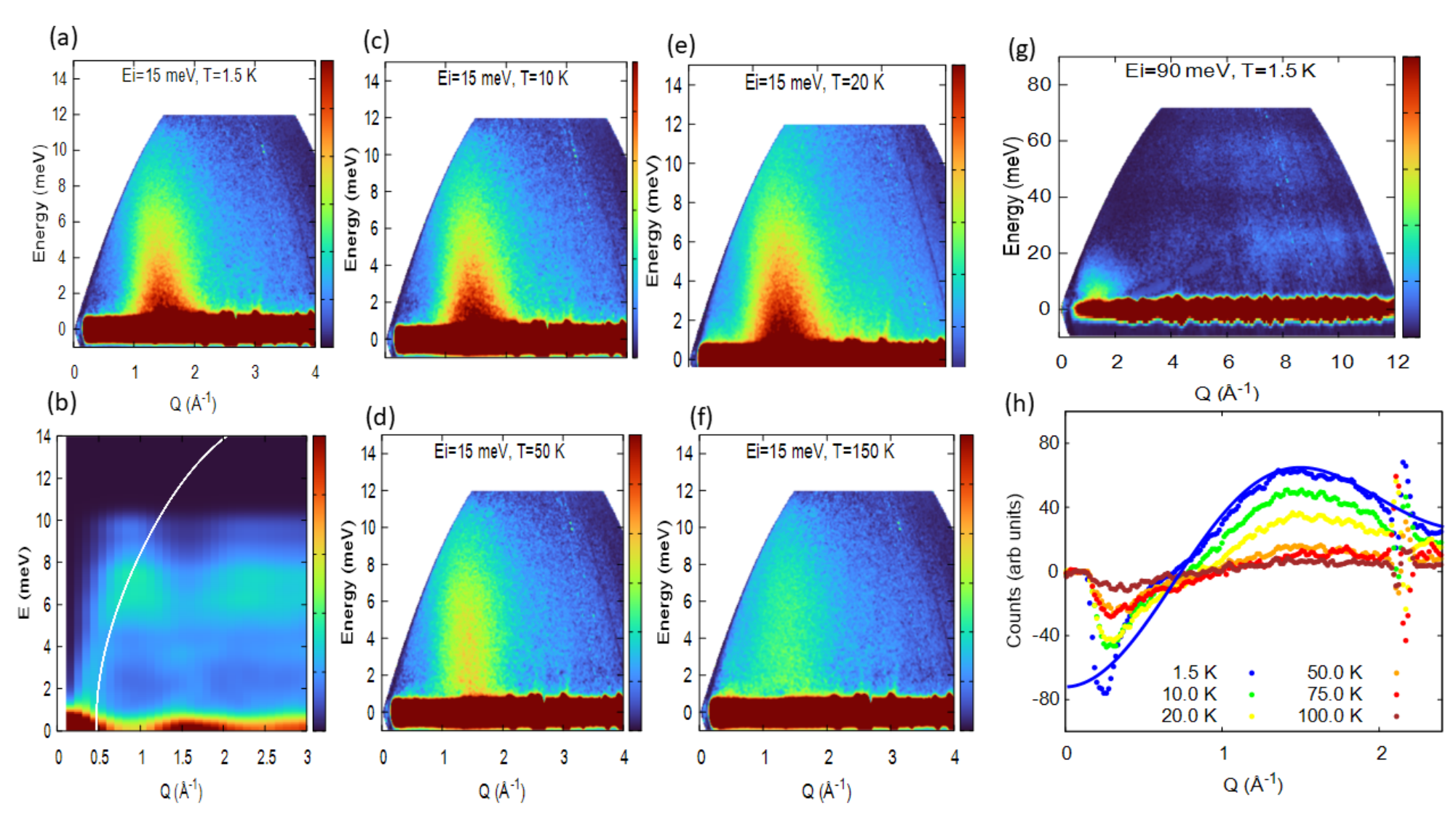}
  \caption{(a) Low-energy INS spectra measured with an incident energy of $E_i = 15$~meV at $T=1.5$ K. (b) The corresponding spin-wave calculated spectra at \( T = 0\,\text{K} \) with an antiferromagnetic exchange interaction \( J = 58\,\text{K} \). (c--f) Low-energy spectra measured with an incident energy of \( E_i = 15\,\text{meV} \) at various temperatures. (g) Time-of-flight (TOF) data recorded with $E_i = 90$~meV at 1.5~K, showing a magnetic signal near $Q = 1.5$~\AA$^{-1}$.  
(h) Energy-integrated ($ E = [-1,1]$~meV), momentum ($Q$) dependent intensity at different temperatures for $E_i = 15$~meV after subtracting the background contribution at $T = 150$~K; the solid line represents a fit using a phenomenological model (see SM~\cite{supplement}), revealing the presence of antiferromagnetic short-range spin correlations.
}
  \label{Neutron}
\end{figure*}

The ZF and LF-$\mu$SR  asymmetries are fitted with a stretched exponential and a temperature-independent background (see SM\cite{supplement}) and the obtained parameters from ZF-$\mu$SR data are summarized in Figs.~\ref{fig:stacked}(b) and \ref{fig:stacked}(c). In the high-$T$ paramagnetic phase, $\lambda_\textrm{ZF}$ shows a slight increment with decreasing temperature, reflecting weakly correlated spins for $T>4$~K and a relaxation mechanism predominantly governed by exchange fluctuations between the Cr moments. Using the previously determined exchange coupling constant \( J = 58 \, \textrm{K} \), the exchange fluctuation rate is \( \nu = \sqrt{z}JS/\hbar \sim 11.67 \times 10^{12} \, \textrm{s}^{-1} \). The relation $\lambda_\textrm{ZF}=2\Delta^2/\nu$ in the motional narrowing limit, $\lambda_\textrm{ZF}(T=20~\textrm{K})=0.098$~$\mu$s$^{-1}$ provides a field distribution width $\Delta\sim8.8~\textrm{kG}~(\sim756$~MHz), apparently smaller than the exchange fluctuation rate $\nu$.

As the temperature is further lowered, $\lambda_\textrm{ZF}(T)$ manifests a steep increase through 3~K and subsequent temperature-independent behavior below 1.3~K. The former signifies the slowing down of magnetic fluctuations by the onset of short-range spin correlations between the Cr moments. The latter is ascribed to persistent spin dynamics, widely reported in frustrated spin systems hosting a spin liquid, weak magnetic order, and spin freezing~\cite{PhysRevLett.98.077204, PhysRevB.91.104427, Cai_2018, PhysRevB.101.224420}. The anomalous low-$T$ behavior of $\lambda_\textrm{ZF}(T)$ is known to arise from unconventional low-energy excitations, though its origin remains unclear~\cite{PhysRevB.91.104427, PhysRevB.54.9019}. Therefore, the observed persistent spin dynamics reflects the presence of unconventional low-lying excitations in the ground state.

The LF dependence of $\mu$SR spectra at 80~mK is illustrated in Fig.~\ref{fig:stacked}(d). As the LF increases, the tail of the muon asymmetry begins to recover gradually. However, the applied LF of 3~kG is not sufficient to fully decouple muons from internal fields, suggesting dynamical fluctuating moments in the ground state. Figure~\ref{fig:stacked}(e) shows the LF dependence of the muon spin relaxation rate. In the low field regime ($H_\textrm{LF}<100$~G), we observe a slight increase in $\lambda_\textrm{LF}$, indicating the decoupling of muon spins from nuclear dipolar fields.. Upon further increasing, $\lambda_\textrm{LF}$ decreases abruptly above 1~kG. The field evolution of the muon spin relaxation rate is well described by the Redfield formalism, $\lambda_\textrm{LF}(H_\textrm{LF})=2\gamma_\mu^2\langle H_\textrm{loc}^2\rangle\nu/(\nu^2+\gamma_\mu^2H_\textrm{LF}^2)$. Here, $\nu$ is the fluctuation rate and $\langle H_\textrm{loc}\rangle$ is the time-average of time-varying local fields of electronic origin at muon sites. The best results are obtained with the parameters $\nu=296(8)$~MHz and $\langle H_\textrm{loc} \rangle=118(1)$~G, as depicted by the solid curve in Fig.~\ref{fig:stacked}(e).  The extracted values are between $\nu\sim150$ MHz and $\langle H_\textrm{loc} \rangle\sim18$~G for herbertsmithite~\cite{PhysRevLett.98.077204} and $\nu\sim1600$ MHz and $\langle H_\textrm{loc} \rangle\sim296$~G for disordered breathing pyrochlore~\cite{lee2021dichotomy}.

It is noteworthy that the LF dependence of the stretched exponent for MCGO is anomalous compared with the typical expected behavior for quantum spin liquid-like systems [Fig.~\ref{fig:stacked}(f)]. $\beta_\textrm{LF}$ generally exhibits an increment and saturation to 1 with increasing field because of quenching of inhomogeneous local fields by external field. Thus, the observed anomalous LF dependence of $\beta_\textrm{LF}$ signifies the field modulation of relaxation channels, possibly reflecting the large distribution of exchange interactions with subtly different energy scales due to the coexistence of fragmented spin clusters and long-range connected spins. Despite the unavoidable site sharing between Cr and Mg, the pyrochlore-like frustrated quantum material MCGO exhibits no signs of spin freezing down to 80 mK. Furthermore, the current 3D spin lattice with substantial exchange interaction maintains a dynamic ground state down to 57 mK, presenting a rare realization in frustrated magnets with quenched disorder. We stress that \(\chi(T) \sim T^{-0.6}\) and \(C_m \sim T^{2.2}\) signatures are incompatible with a simple random singlet scenario.

\noindent\hspace{1em}
To shed insights into low-energy excitations and spin correlations in the 3D spin liquid candidate MCGO, INS experiments were conducted down to 1.5~K at various incident energies, as shown in Fig.~\ref{Neutron}. The high-temperature spectra are dominated by phonon density of states; however, a rod-like feature emerges at $Q = 1.5~\text{\AA}^{-1}$, whose intensity increases upon decreasing temperature below 20~K (Figs.~\ref{Neutron}(a)), most likely of magnetic origin (see SM~\cite{supplement}). The spin-wave calculations capture the rise of diffuse scattering close to \( 1.5\,\text{\AA}^{-1} \) (Fig.~\ref{Neutron}(b)) with a nearest-neighbor antiferromagnetic interaction, \( J = 58\,\text{K} \). Despite the broadening of the spectra due to the combination of disorder and powder averaging, it remains feasible to identify acoustic-like modes stemming from \( 1.5\,\text{\AA}^{-1} \), akin to experimental results. The spectra exhibit broad features extending up to an energy that roughly scales with \( J \) (see SM~\cite{supplement}). Strikingly, the quasi-elastic line, defined by an integration over energies $ E = [-1,1]$~meV, reveals a broad peak centered at $Q = 1.5~\text{\AA}^{-1}$, implying the emergence of strong magnetic diffuse scattering at this wavevector at low temperatures. The absence of magnetic Bragg peaks in the INS spectra rules out any long-range magnetic ordering in this 3D frustrated antiferromagnet. The quasi-elastic line is further analyzed after subtracting the background contribution measured at $T = 150$~K, as presented in Fig.~~\ref{Neutron}(h). The observed $Q$-dependence of the magnetic diffuse scattering at low temperatures can be well captured by a phenomenological model (see SM~\cite{supplement}), which yields a correlation length of approximately 3~\AA, corresponding to the nearest-neighbor distance between Cr$^{3+}$ ($S=3/2$) moments in the host pyrochlore spin lattice. This implies the presence of antiferromagnetic short-range spin correlations, a typical feature of spin liquid candidates. INS experiments do not detect a gap in the excitation spectra. The enhancement of the intensity of the quasi-elastic magnetic diffuse scattering upon cooling indicates the development of antiferromagnetic short-range spin correlations and a slowing down of spin dynamics, consistent with complementary thermodynamic, ESR, and $\mu$SR results.


\noindent\hspace{1em} 
To summarize, the unavoidable anti-site disorder in highly frustrated pyrochlore MCGO introduces spatially varying magnetic correlations mediated by exchange randomness, suppressing long-range magnetic order. Thermodynamic experiments detect no signature of long-range magnetic ordering or spin freezing down to 57 mK. The $\mu$SR experiment confirms the absence of magnetic order and spin freezing, establishing a persistent dynamic state down to 80 mK. The ESR measurements indicate
the onset of short-range spin correlations below 15 K, which is corroborated by the presence of a broad maximum at $\sim$5 K  in specific heat. Furthermore, low-energy inelastic scattering uncovers magnetic diffuse scattering centered on $Q = 1.5$\,\AA$^{-1}$ at temperature $T \sim J$, implying the development of antiferromagnetic short-range spin correlations, which are quasi-static at the typical time scale of neutron experiments and reinforcing a putative spin liquid state in this three-dimensional frustrated magnet. The power-law behavior of magnetic specific heat below 1 K, that is supported by the absence of a gap in the INS experiment, suggests algebraic spin correlations in the classical spin liquid state of this three-dimensional disordered pyrochlore-like frustrated magnet. The present study may instigate further efforts for the realization of elusive quantum states with exotic excitations in higher-dimensional frustrated lattices.

\noindent\hspace{1em} \textit{Acknowledgements.}
P.K. acknowledges the funding by the Science and Engineering Research Board, and Department of Science and Technology, India through Research Grants. This work was supported by the Institute for Basic Science (IBS-R014-Y2). The authors thank S. Yoon of the National Center for Inter-university Research Facilities at Seoul National University for assistance with the X-band ESR experiments. This work at SKKU was supported by the National Research Foundation
(NRF) of Korea (Grants No. RS-2023–00209121
and No. 2022R1A2C1003959).

\noindent\hspace{1em} \textit{Data availability.}
The data that support the findings of the current study are available from the corresponding author upon reasonable request.
\bibliography{MCGO}

\end{document}